\documentclass{ws-ijmpe}

\usepackage{graphicx}
\begin{document}

\markboth{Authors' Names}{The Prompt Dipole Radiation}
\catchline{}{}{}{}{}

\title{THE DYNAMICAL DIPOLE RADIATION IN DISSIPATIVE COLLISIONS WITH EXOTIC 
BEAMS}
\author{M. DI TORO $^*$, M. COLONNA, C. RIZZO}%
\address{Laboratori Nazionali del Sud INFN, I-95123 
Catania, Italy,\\
and Physics-Astronomy Dept., University of Catania\\%
$^*$ E-mail:\_ditoro@lns.infn.it}

\author{V. BARAN}%
\address{Dept.of Theoretical Physics, Bucharest Univ., and NIPNE-HH,
Magurele, Bucharest, Romania}

\maketitle
\begin{history}%
\received{September 07}%
\revised{September 07}%
\end{history}

\begin{abstract}
Heavy Ion Collisions ($HIC$) represent a unique tool to probe the in-medium
nuclear interaction in regions away from saturation. In this work we present a 
selection of reaction observables in dissipative collisions 
particularly sensitive to the isovector part of the interaction, i.e. to the
symmetry term of the nuclear Equation of State ($EoS$).
At low energies the behavior of the symmetry energy around saturation 
influences dissipation and fragment production mechanisms.
We will first discuss the recently observed Dynamical Dipole Radiation, due
to a collective neutron-proton oscillation during the charge equilibration
in fusion and deep-inelastic collisions. We will review in detail all the main
properties, yield, spectrum, damping and angular distributions, revealing 
important isospin effects. Reactions induced by unstable $^{132}Sn$ beams appear
to be very promising tools to test the sub-saturation Isovector $EoS$.
Predictions 
are also presented for deep-inelastic and fragmentation collisions induced by neutron 
rich projectiles. The importance of studying violent collisions with radioactive beams
at low and Fermi energies is finally stressed.

\end{abstract}


\vskip -1.0cm
\section{Introduction}

The symmetry energy $E_{sym}$ appears in the energy density
$\epsilon(\rho,\rho_3) \equiv \epsilon(\rho)+\rho E_{sym} (\rho_3/\rho)^2
 + O(\rho_3/\rho)^4 +..$, expressed in terms of total ($\rho=\rho_p+\rho_n$)
 and isospin ($\rho_3=\rho_p-\rho_n$) densities. The symmetry term gets a
kinetic contribution directly from basic Pauli correlations and a potential
part from the highly controversial isospin dependence of the effective 
interactions \cite{baranPR}. Both at sub-saturation and supra-saturation
densities, predictions based of the existing many-body techniques diverge 
rather widely, see \cite{fuchswci}. 
We take advantage of new opportunities in 
theory (development of rather reliable microscopic transport codes for $HIC$)
 and in experiments (availability of very asymmetric radioactive beams, 
improved possibility of measuring event-by-event correlations) to present
results that are severely constraining the existing effective interaction 
models. We will discuss dissipative collisions in the low energy range, 
 from just above the Coulomb barrier up to about hundred $AMeV$. In this 
way we can probe in detail the symmetry energy in dilute matter. 
The transport codes are based on 
mean field theories, with correlations included via hard nucleon-nucleon
elastic and inelastic collisions and via stochastic forces, selfconsistently
evaluated from the mean phase-space trajectory, see 
\cite{baranPR,guarneraPLB373,colonnaNPA642,chomazPR}. Stochasticity is 
essential in 
order to get distributions as well as to allow the growth of dynamical 
instabilities.  
 The isovector part of the $EoS$ has been tested systematically by using two 
different behaviors of the symmetry energy below saturation: 
one ($Asysoft$) where it is a smooth decreasing function towards low densities,
 and another one ($Asystiff$) where we have a rapid decrease, 
\cite{baranPR,colonnaPRC57}.

\section{The Prompt Dipole $\gamma$-Ray Emission}

The possibility of an entrance channel bremsstrahlung dipole radiation
due to an initial different N/Z distribution was suggested at the beginning
of the nineties \cite{ChomazNPA563,BortignonNPA583}, largely inspired by
D.M.Brink discussions. At that time a large debate was present on the 
disappearing
of $Hot~Giant~Dipole~Resonances$ in fusion reactions. Brink was suggesting
the simple argument that a $GDR$ needs time to be built in a hot compound
 nucleus, meanwhile the system will cool down by neutron emission and the
$GDR$ photons will show up at lower temperature. The natural consequence
suggested in ref. \cite{ChomazNPA563} was that we would expect a new
dipole emission, in addition to the statistical one, if some pre-compound
collective dipole mode is present.
After several experimental evidences, in fusion as well as in deep-inelastic
reactions \cite{FlibPRL77,CinNC111,PierrouEPJA16,AmoPRC29,PierrouPRC71} we have
now a good understanding of the process and stimulating new perspectives
from the use of radioactive beams.

During the charge equilibration process taking place
 in the first stages of dissipative reactions between colliding ions with
 different N/Z
ratios, a large amplitude dipole collective motion develops in the composite
dinuclear system, the so-called dynamical dipole mode. This collective dipole
gives rise to a prompt $\gamma $-ray emission which depends:
 i) on the absolute
value of the intial dipole moment
\begin{eqnarray}
&&D(t= 0)= \frac{NZ}{A} \left|{R_{Z}}(t=0)- {R_{N}}(t=0)\right| =  \nonumber \\
&&\frac{R_{P}+R_{T}}{A}Z_{P}Z_{T}\left| (\frac{N}{Z})_{T}-(\frac{N}{Z})_{P}
\right|,
\label{indip}
\end{eqnarray}
being ${R_{Z}}= \frac {\Sigma_i x_i(p)}{Z}$ and
${R_{N}}=\frac {\Sigma_i x_i(n)}{N} $ the
center of mass of protons and of neutrons respectively, while R$_{P}$ and
R$_{T}$ are the
projectile and target radii; ii) on the fusion/deep-inelastic dynamics;
 iii) on the symmetry term, below saturation, that is acting as a restoring
force.

A detailed description is obtained in a
microscopic approach based on semiclassical transport equations,
of Landau-Vlasov type, \cite{BrinkNPA372},
where mean field and two-body collisions are treated in a
self-consistent way, see details in \cite{BaranNPA600}. Realistic
effective interactions of Skyrme type are used. The numerical
accuracy of the transport code has been largely improved in order
to have reliable results also at low energies, just above the
threshold for fusion reactions
\cite{CabNPA637,BaranNPA679}. The
resulting physical picture is in good agreement with quantum
Time-Dependent-Hartree-Fock calculation \cite{SimenPRL86}. In
particular we can study in detail how a collective dipole
oscillation develops in the entrance channel \cite{BaranNPA679}.

First, during the ${\it approaching~phase}$, the two partners,
overcoming the Coulomb barrier, still keep their own response.
Then a {\it dinuclear phase} follows, where the
relative motion energy, due to the nucleon exchange, is converted
in thermal motion and in
the collective
energy of the dinuclear mean field.
In fact the composite system is not
fully  equilibrated and  manifests, as a whole, a large
amplitude  dipole collective motion. Finally thermally
equilibrated reaction products are formed, with consequent
statistical particle/radiation emissions.

\begin{figure}
\begin{center}
\includegraphics[scale=0.37]{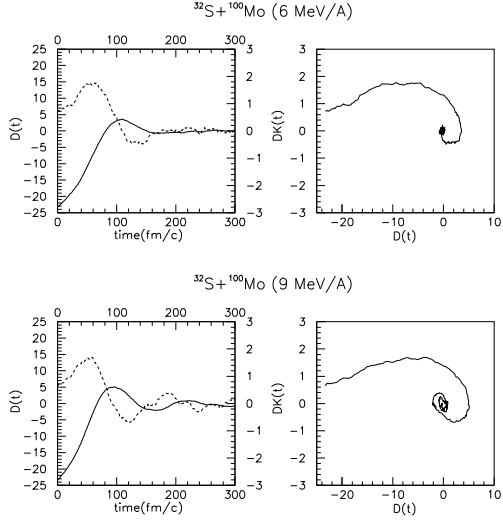}
\caption{
The time evolution of the dipole mode
in $r-$space $D(t)$ (solid lines) and $p-$space $DK(t)
$
(dashed lines, in
$fm^{-1}$) and the correlation $DK(t) -  D(t)$
at incident energy of $6AMeV$ and $9AMeV$ for $b= 2 fm$.}
\end{center}
\label{spiral}
\end{figure}

\begin{figure}
\begin{center}
\includegraphics[scale=0.40]{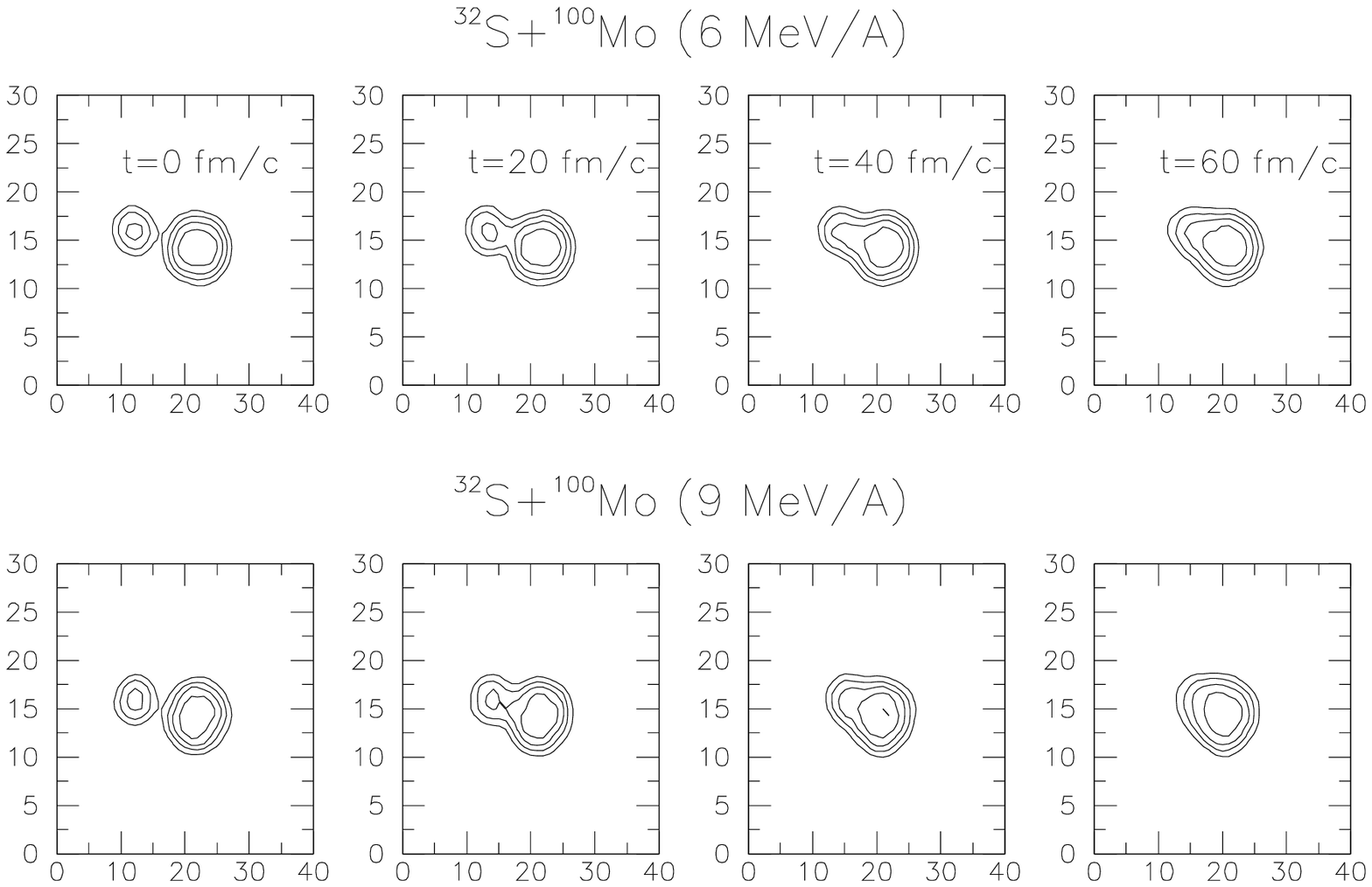}
\caption{Density plots of the neck dynamics for the $^{32}$S +$^{100}$Mo
system at incident energy of $6AMeV$ and $9AMeV$.}
\end{center}
\label{fusdyn}
\end{figure}

We present here some results for the $^{32}$S +$^{100}$Mo ($N/Z$ asymmetric)
reaction at $6~and~9~A MeV$, recently studied vs. the
``symmetric'' $^{36}$S +$^{96}$Mo
counterpart in ref.\cite{PierrouPRC71}.
In Fig.1 (left columns) we draw the time
evolution for $b= 2 fm$ of the dipole moment
in the $r$-space (solid lines),
 $D(t)= \frac{NZ}{A} ({R_{Z}}- {R_{N}})$ and in
$p-$space (dashed lines), $DK(t)=(\frac{P_{p}}{Z}-\frac{P_{n}}{N})$, with $P_{p}$
($P_{n}$) center of mass in momentum space for protons (neutrons),
is just the canonically conjugate momentum of the $D(t)$ coordinate,
i.e. as operators $[D(t),DK(t)]=i\hbar$
see \cite{BaranNPA679,SimenPRL86,BaranPRL87}. On the right hand side
columns we show the
corresponding correlation $DK(t) -  D(t)$
in the phase space.
We choose the origin of time at the
beginning of the {\it dinuclear} phase. The nice "spiral-correlation"
clearly denotes the collective nature
 of the mode.
>From Fig.\ref{spiral} we note that the "spiral-correlation" starts when
the initial dipole moment $D(t= 0)$, the geometrical value at the
touching point, is already largely quenched. This is the reason
why the dinucleus dipole yield is not simply given by the "static"
estimation but the reaction dynamics has a large influence on it.

A clear energy
dependence of the dynamical dipole mode is evidenced
with a net increase when we pass
from $6AMeV$ to $9AMeV$. A possible explanation
of this effect is due to the fact that at lower energy, just above the Coulomb
barrier, a longer
transition to a dinuclear configuration is required
which hinders the isovector
collective response. From Fig.\ref{fusdyn}
a  slower dynamics of the neck
during the first $40fm/c - 60fm/c$ from the touching
configuration is observed at $6~AMeV$. When the collective dipole response
sets in the charge is already partially equilibrated via random nucleon
exchange.

The bremsstrahlung spectra shown in Fig.3 support this
interpretation.

\begin{figure}
\begin{center}
\includegraphics[scale=0.37]{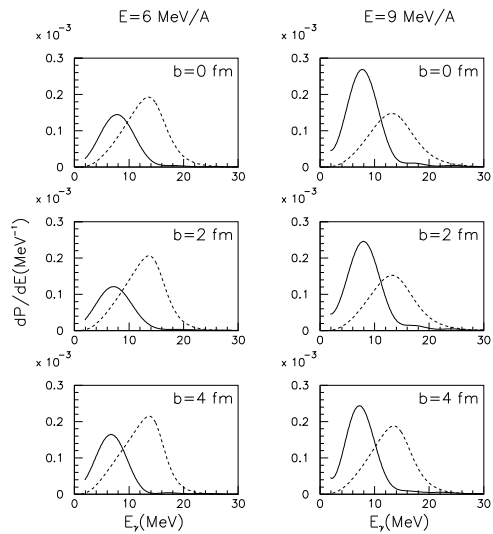}
\caption{The bremsstrahlung spectra for the $^{32}$S +$^{100}$Mo system
at incident energy of $6AMeV$ and $9AMeV$ (solid line) and the first step
statistical spectrum (dashed line) for three
impact parameters.}
\end{center}
\label{spectra}
\end{figure}

In fact from the dipole evolution given from the Landau-Vlasov transport
we can directly
apply a bremsstrahlung {\it ("bremss")}  approach
\cite{BaranPRL87} to estimate the ``direct'' photon emission probability
($E_{\gamma}= \hbar \omega$):
\begin{equation}
\frac{dP}{dE_{\gamma}}= \frac{2 e^2}{3\pi \hbar c^3 E_{\gamma}}
 |D''(\omega)|^{2}  \label{brems},
\end{equation}
where $D''(\omega)$ is the Fourier transform of the dipole acceleration
$D''(t)$. We remark that in this way it is possible
to evaluate, in {\it absolute} values, the corresponding pre-equilibrium
photon emission.
In the same Fig.3 we show statistical $GDR$ emissions from the
final excited residue. We see that at the higher energy the prompt emission
represents a large fraction of the total dipole radiation.

\begin{table*}[hbt]
\caption{\it The percent increase of the intensity in the linearized
$\gamma $-ray
spectra at the compound nucleus GDR energy region (the energy integration
 limits are given in the parenthesis), the compound
nucleus excitation energy, the initial dipole moment
$D(t=0)$ and the initial mass asymmetry $\Delta$ for each reaction.}
\begin{tabular}{lcccccccccc}
Reaction  &  Increase (\%) & E$^{*}$(MeV)& $D(t=0)$ (fm) & $\Delta$ &
 Ref \\
\hline
$^{40}$Ca+$^{100}$Mo & 16 (8,18)& 71 &22.1 & 0.15 & \cite{FlibPRL77} \\
$^{36}$S+$^{104}$Pd & & 71 & 0.5 & 0.17   \\
\hline
$^{16}$O+$^{98}$Mo &  36 (8,20)& 110 &8.4 & 0.29 & \cite{CinNC111} \\
$^{48}$Ti +$^{64}$Ni & & 110 & 5.2 & 0.05 \\
\hline
$^{32}$S+$^{100}$Mo &  1.6 $\pm 2.0$ (8,21)& 117 &18.2 & 0.19 &
\cite{PierrouPRC71}\\
$^{36}$S+$^{96}$Mo &  & 117 &1.7 & 0.16 \\
\hline
$^{32}$S+$^{100}$Mo &  18.0 $\pm 2.0$ (8,21)& 173.5 &18.2 & 0.19 & \cite{PierrouEPJA16} \\
$^{36}$S+$^{96}$Mo &  & 173.5 &1.7 & 0.16 \\
\hline
$^{36}$Ar+$^{96}$Zr &  12.0 $\pm 2.0$ (8,21)& 280  & 20.6 & 0.16 & \cite{Medea} \\
$^{40}$Ar+$^{92}$Zr &  & 280 &4.0 & 0.14 
\end{tabular}
\end{table*}

In the Table we report the present status of the Dynamical Dipole data,
obtained from fusion reactions.
We note the dependence of the extra strength on the interplay between
initial dipole moment and initial mass asymmetry \cite{deltadef}: this clearly indicates
the relevance of the fusion dynamics.

We must add a
couple of comments of interest for the experimental selection of the Dynamical
Dipole: i) The centroid is always shifted to lower energies (large
deformation of the dinucleus, slightly increasing with impact parameter); 
ii) A clear angular anisotropy should be present
since the prompt mode has a definite axis of oscillation
(on the reaction plane) at variance with the statistical $GDR$.

In a very recent experiment the prompt dipole radiation has been investigated with
a $4 \pi$ gamma detector. A strong dipole-like photon angular distribution
$(\theta_\gamma)=W_0[1+a_2P_2(cos \theta_\gamma)]$, $\theta_\gamma$ being the 
angle between the emitted photon and the beam axis, has been observed, with the 
$a_2$ parameter close to $-1$, see \cite{Medea}. The deviation from a $pure$
dipole form can be interpreted as due to the rotation of the dinucleus symmetry
axis vs. the beam axis during the Prompt Dipole Emission. From accurate angular 
distribution mesurements we can then expect to get a direct information on the
Dynamical Dipole Life Time.

At higher beam energies we expect a further decrease of the direct dipole
radiation for two main reasons both due to the increasing importance of hard
$NN$ collisions: i) a larger fast nucleon emission that will equilibrate the
isospin before the collective dipole starts up; ii) a larger damping of the
collective mode due to $np$ collisions. This has been observed in 
ref.\cite{AmoPRC29} and more exps.
are planned \cite{Medea}.

The prompt dipole radiation also represents a nice cooling mechanism on the
fusion path. It could be a way to pass from a {\it warm} to a {\it cold}
fusion in the synthesis of heavy elements with a noticeable increase of the
{\it survival} probability, \cite{luca04},

\begin{equation}
\frac{P_{surv,dipole}}{P_{surv}}=\frac{P_\gamma P_{surv}(E^*-E_\gamma)}
{P_{surv}(E^*)} + (1-P_\gamma) > 1, \frac{P_{surv}(E^*-E_\gamma)}{P_{surv}(E^*)} > 1.
\label{surv}
\end{equation}

\begin{figure}[htb]
\begin{center}
\includegraphics[scale=0.25]{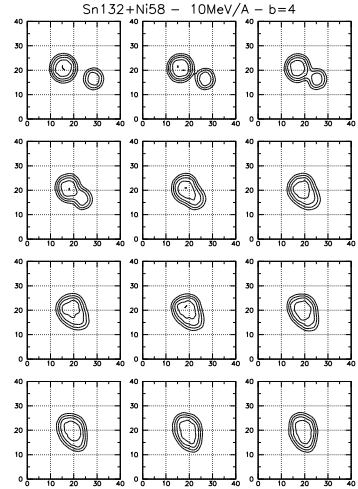}
\caption{$^{132}$Sn +$^{58}$Ni system ($E=10 AMeV$, $b=4fm$).
Density contour plots (20fm/c time steps)}
\end{center}
\label{contours}
\end{figure}

\vskip -1.0cm
\subsection{Symmetry Energy Effects}

The use of unstable neutron rich projectiles would largely increase the
effect, due to the possibility of larger entrance channel asymmetries.
\begin{figure}[htb]
\begin{center}
\centering
\includegraphics[width=8.8cm]{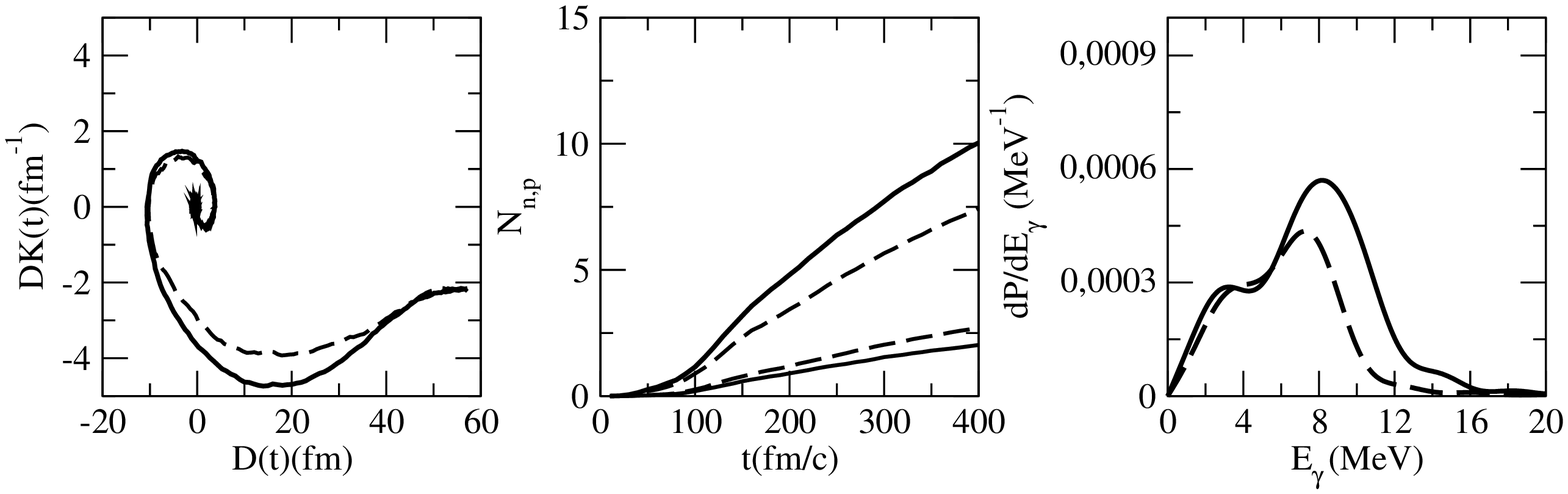}
\vskip 1.8cm
\includegraphics[width=8.8cm]{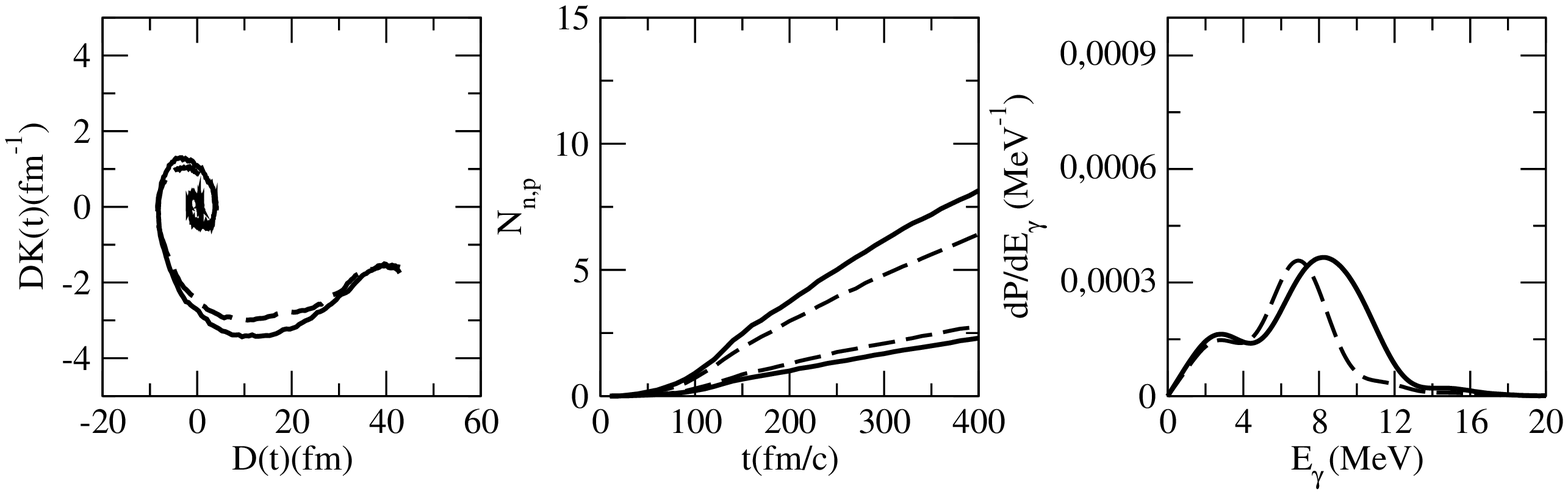}
\caption{Upper Curves: $^{132}$Sn +$^{58}$Ni system ($E=10 AMeV$, $b=4fm$).
 Lower Curves: same reactions but induced by $^{124}$Sn.
Left Panel: DK-D Spirals. Central panel: neutron (upper) and proton (lower) 
emissions. Right panel: $\gamma$ spectrum. Asy-soft (solid lines) and asy-stiff
(dashed lines) symmetry energies.}
\end{center}
\label{sn132}
\end{figure}
In order to suggest proposals for the new $RIB$ facility $Spiral~2$, 
\cite{lewrio} we have studied fusion events in the reaction $^{132}Sn+^{58}Ni$ 
at $10AMeV$, \cite{spiral2}. 
We espect a $Monster$ Dynamical Dipole, the initial 
dipole moment $D(t=0)$ being of the order of 50fm, about two times the larger 
values of the Table, allowing a detailed study of the symmetry potential, below 
saturation,
responsible of the restoring force of the dipole oscillation and even of the damping,
 via the fast neutron emission. 
The corresponding contour plots on the reaction plane are shown in Fig.4.
 We note the clear rotation of the symmetry axis during the prompt dipole emission.
 
In the Fig.5 (upper) we present some preliminary very promising results.
The larger value of the symmetry energy for the $Asysoft$ choice at low densities,
where the prompt dipole oscillation takes place, leads to some clear observable 
effects:
i) Larger Yields, as we see from the larger amplitude of the "Spiral" (left panel) 
and
finally in the spectra (right panel); ii) Larger mean gamma energies, shift of the 
centroid
to higher values in the spectral distribution (right panel); iii) Larger width of the
"resonance" (right panel) due to the larger fast neutron emission (central panel).
We note the opposite effect of the Asy-stiffness on neutron vs proton emissions.
The latter point is important even for the possibility of an independent test just
measuring the $N/Z$ of the pre-equilibrium nucleon emission. 

The symmetry energy influence can be of the order of $20\%$, and so well detected.
In the lower part of the same figure we present the same results for reactions induced by
the stable $^{124}Sn$ beam. We still se the $Iso-EoS$ effects, but largely reduced. 

\section{Isospin effects on Deep-Inelastic Collisions}
Dissipative semi-peripheral collisions at low energies, including binary and 
three-body breakings, offer a good opportunity to study phenomena occurring 
in nuclear matter under extreme conditions with respect to shape, excitation 
energy, spin and N/Z ratio (isospin). 
In some cases, due to a combined Coulomb and angular momentum 
(deformation) effect, some instabilities can show up \cite{colonnaNPA589}.
 This can lead to 3-body breakings, where a light cluster is emitted from the
 neck region. Three body processes in collisions with exotic beams will allow 
to investigate how the development of surface (neck-like) instabilities, that 
would help ternary breakings, is sensitive to the structure of the symmetry 
term around (below) saturation.
In order to suggest proposals for the new $RIB$ facilities, we have studied 
again the reaction $^{132}Sn+^{64}Ni$ at $10AMeV$
in semicentral events, impact parameters $b=6, 7, 8 fm$, where one observes 
mostly binary exit channels, but still in presence of large dissipation.
 
\begin{figure}
\vskip 0.7cm
\begin{center}
\includegraphics[scale=0.35]{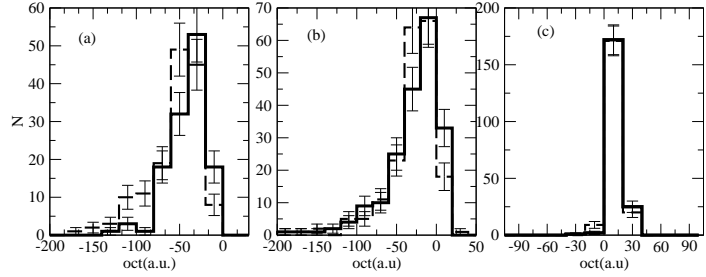}
\caption{Distribution of the octupole moment of primary fragments for 
the $^{132}Sn+^{64}Ni$
reaction at $10~AMeV$ (impact parameters (a):$b=6fm$, (b):$7fm$, (c):$8fm$).
Solid lines: asysoft. Dashed lines: asystiff} 
\label{octupole}
\end{center}
\end{figure}

The Wilczynski plots, kinetic energy loss vs. deflection angle, show slightly 
more dissipative events in the $Asystiff$ case, consistent with the point that 
in the interaction at lower densities in very neutron-rich matter (the neck 
region) we have a less repulsive symmetry term. In fact the 
neck dynamics is 
rather different in the two cases, as it can be well  evidenced looking at the
 deformation of the $PLF/TLF$ residues. The distribution of the octupole 
moment over the considered ensemble of events is shown in Fig.\ref{octupole} 
for the three considered impact parameters.

Except for the most peripheral events, larger deformations, strongly 
suggesting a final 3-body outcome, are seen in the $Asystiff$ case. Now, 
due to the lower value of the symmetry enrgy, the neutron-rich neck 
connecting the two systems survives a longer time leading to very deformed 
primary fragments, from which eventually small clusters will be dynamically 
emitted.
Finally we expect to see effects of the different interaction times on the 
charge equilibration mechanism, probed starting from entrance channels with 
large $N/Z$ asymmetries, like $^{132}Sn(N/Z=1.64)+^{58}Ni(N/Z=1.07)$. Moreover 
the equilibration mechanism is also directly driven by the strenght of the 
symmetry term.

\vskip -1.0cm
\section{Isospin Dynamics in Neck Fragmentation at Fermi Energies}

It is now quite well established that the largest part of the reaction
cross section for dissipative collisions at Fermi energies goes
through the {\it Neck Fragmentation} channel, with $IMF$s directly
produced in the interacting zone in semiperipheral collisions on very short
time scales \cite{wcineck}. We can predict interesting isospin transport 
effects for this new
fragmentation mechanism since clusters are formed still in a dilute
asymmetric matter but always in contact with the regions of the
projectile-like and target-like remnants almost at normal densities.
Since the difference between local neutron-proton chemical potentials is given 
by $\mu_n-\mu_p=4E_{sym}(\rho_3/\rho)$, and the isospin transport is ruled 
by the density gradient,  we expect a larger neutron flow to
 the neck clusters for a stiffer symmetry energy around saturation, 
\cite{baranPR,baranPRC72}. The isospin dynamics can be directly extracted 
from correlations between $N/Z$, $alignement$ and emission times of the $IMF$s.
The alignment between $PLF-IMF$ and $PLF-TLF$ directions
represents a very convincing evidence of the dynamical origin of the 
mid-rapidity fragments produced on short time scales \cite{baranNPA730}. 
The form of the
$\Phi_{plane}$ distributions (centroid and width) can give a direct
information on the fragmentation mechanism \cite{dynfiss05}. Recent 
calculations confirm that the light fragments are emitted first, a general 
feature expected for that rupture mechanism \cite{liontiPLB625}. 
The same conclusion can be derived from {\it direct} emission time 
measurements based on deviations from Viola systematics  observed
in event-by-event velocity correlations between $IMF$s and the $PLF/TLF$ 
residues
 \cite{baranNPA730,dynfiss05,velcorr04}. 
 We can figure out
   a continuous transition from fast produced fragments via neck instabilities
   to clusters formed in a dynamical fission of the projectile(target) 
   residues up to the evaporated ones (statistical fission). Along this 
   line it would be even possible to disentangle the effects of volume
   and shape instabilities. 
A neutron enrichment of the overlap ("neck") region is
   expected, due to the neutron migration from higher (spectator) to 
   lower (neck) density regions, directly related to
   the slope of the symmetry energy \cite{liontiPLB625}. 
A very nice new analysis has been presented on the $Sn+Ni$ data at $35~AMeV$
by the Chimera Collab., Fig.2 of ref.\cite{defilposter}.
A strong correlation between neutron enrichemnt and alignement (when the 
short emission time selection is enforced) is seen, that can be reproduced 
only with 
a stiff behavior of the symmetry energy. {\it This is the 
first clear evidence in favor of a relatively large slope (symmetry pressure) 
around saturation}.

\vskip -1.0cm
\section{Perspectives}
We have shown that {\it violent} collisions of n-rich heavy ions 
from low to Fermi energies
can bring new information on the isovector part of the in-medium interaction, 
qualitatively different from equilibrium
$EoS$ properties, in particular in dilute nuclear matter. 
We have presented quantitative results 
suggesting several isospin-sensitive 
observables.
At low energies we have seen isospin effects on the rather exciting new
prompt dipole radiation and on the dissipation in deep 
inelastic collisions, at Fermi energies the Iso-EoS sensitivity of the isospin 
transport in fragment reactions. 

In conclusion the results presented here appear very promising for 
the possibility of exciting new results from dissipative collisions
with radioactive beams.
                                                                      %



\end{document}